\documentclass[journal]{IEEEtran}
\usepackage{cite}
\usepackage{amsmath}
\usepackage{graphicx}
\ifCLASSINFOpdf

\else

\fi

\begin{document}
	%
	\title{Filter Drug-induced Liver Injury Literature with Natural Language Processing and Ensemble Learning}
	%
	%
	%
	
	\author{Xianghao Zhan, Fanjin Wang, Olivier Gevaert \thanks{X. Zhan is with the Department of Bioengineering, Stanford University,
			Stanford, 94305, USA. 
			
			F. Wang is with the Department of Pharmaceutics, UCL School of Pharmacy, University College London, 29-39 Brunswick Square, London WC1N 1AX, UK.
			
			O. Gevaert is with the Department of Biomedical Informatics, Stanford
			University, Stanford, 94305, USA
			
			(Corresponding Author: O. Gevaert e-mail: ogevaert@stanford.edu)}}%
	
	%
	%
	
	\markboth{}%
	{Shell \MakeLowercase{\textit{et al.}}: Bare Demo of IEEEtran.cls for IEEE Journals}
	%



	\maketitle
	
	\begin{abstract}
		Drug-induced liver injury (DILI) describes the adverse effects of drugs that damage liver. Life-threatening results including liver failure or death were also reported in severe DILI cases. Therefore, DILI-related events are strictly monitored for all approved drugs and the liver toxicity became important assessments for new drug candidates. These DILI-related reports are documented in hospital records, in clinical trial results, and also in research papers that contain preliminary \textit{in vitro} and \textit{in vivo} experiments. Conventionally, data extraction from previous publications relies heavily on resource-demanding manual labelling, which considerably decreased the efficiency of the information extraction process. The recent development of artificial intelligence, particularly, the rise of natural language processing (NLP) techniques, enabled the automatic processing of biomedical texts. In this study, based on around 28,000 papers (titles and abstracts) provided by the Critical Assessment of Massive Data Analysis (CAMDA) challenge, we benchmarked model performances on filtering out DILI literature. Among four word vectorization techniques, the model using term frequency-inverse document frequency (TF-IDF) and logistic regression outperformed others with an accuracy of 0.957 with our in-house test set. Furthermore, an ensemble model with similar overall performances was implemented and was fine-tuned to lower the false-negative cases to avoid neglecting potential DILI reports. The ensemble model achieved a high accuracy of 0.954 and an F1 score of 0.955 in the hold-out validation data provided by the CAMDA committee. Moreover, important words in positive/negative predictions were identified \textit{via} model interpretation. Overall, the ensemble model reached satisfactory classification results, which can be further used by researchers to rapidly filter DILI-related literature.
	\end{abstract}
	
	\begin{IEEEkeywords}
		drug-induced liver injury, natural language processing, ensemble learning, word2vec, sentence embedding
	\end{IEEEkeywords}

	%
	\IEEEpeerreviewmaketitle

	\section{Introduction}
	
	\IEEEPARstart{D}{urg}-induced liver injury (DILI) is defined as the unexpected adverse reaction of the liver to drugs. DILI is a common and critical cause of liver injury because the liver plays a key role in drug metabolism \cite{lee_drug-induced_2003,andrade_drug-induced_2019}. Briefly, liver toxicity caused by drugs can be classified into two types: intrinsic and idiosyncratic. Intrinsic liver toxicity of drugs is more predictable and is directly related to the dosage of a specific drug. The damage to liver occurs within a short time window, typically several hours after administration of the drugs. In comparison, idiosyncratic liver toxicity is more patient-specific and has a longer onset of occurrence. For drugs with high lipophilicity, idiosyncratic liver damage could be triggered even below the recommended daily dosage \cite{chen_high_2013}. The severity of DILI can be different among different patients considering the interaction of genetic and environmental factors \cite{andrade_drug-induced_2019}. Although most patients can recover from DILI, DILI cases may lead to acute liver failure \cite{lee_drug-induced_2003,kaplowitz_drug-induced_2004}. For example, the intrinsic liver toxicity of paracetamol, often caused by overdosing, is reported to account for 73.7\% of acute liver injury and acute liver failure in Scotland from 1992 to 2014 \cite{donnelly_acute_2017}. Additionally, approximately 75\% of the idiosyncratic drug reactions result in liver transplantation or death \cite{lee_drug-induced_2003}. Therefore, DILI has become one of the most common reasons that reject the promising novel drug candidates and is strictly evaluated during the drug development process \cite{andrade_drug-induced_2019}.

	The complex mechanism of DILI and the severity of the DILI consequences call for a better prediction of DILI \cite{andrade_drug-induced_2019}. However, the majority of DILI results from clinical practices or experimental studies have been reported and recorded in the free text of publications. Conventionally, scientific publications need to be manually checked and processed by researchers for these DILI-related results. However, thousands of new articles are published in journals on a daily basis, let alone millions of previous publications in the PubMed archive. This makes it almost impossible for manual inspection. Recently, the rapid development of natural language processing (NLP) technology has enabled data mining applications based on free text. To give some examples, long short-term memory (LSTM) structure in recurrent neural networks (RNN) allows the understanding of long-term dependencies in texts \cite{hochreiter_long_1997}. Bidirectional encoder representation from transformers (BERT) has been developed as a pre-trained language model for understanding text information \cite{devlin_bert_2019}. Generally, in order to use these learning algorithms, words will be firstly converted into vectors using word vectorization (embedding) techniques like bag-of-words (BOW), term frequency-inverse document frequency (TF-IDF), and Word2Vec. BOW is the most straightforward word embedding method. It counts the time that a word appears in the document. However, this leads to a very sparse feature matrix as only a few words in the whole vocabulary (the collection of words) will appear in the document. As its name suggests, TF-IDF uses the time frequency and the inverse document frequency to represent a word. It assigns more importance to less-frequently occurring words which might contain more meaningful information in a document. On top of these methods, Word2Vec embedding uses a pre-trained neural network to vectorize words to fixed-length vectors. These techniques made it possible to process scientific publications automatically. For example, Wang et al. constructed an NLP model to extract clinical information to support clinical decisions \cite{wang_ensemble_2016}. Zhan et al. used TF-IDF as the word embedding and logistic regression to extract ICD-10 codes of common cardiovascular disease from electronic health records \cite{zhan_structuring_2021}. Thus, it is promising to utilize NLP techniques to expedite the labelling process of the publications with DILI results and enable researchers to fast filter the literature.

	In this study with the data from the National Institute of Health (NIH) LiverTox database \cite{hoofnagle_livertox_2013}, we developed a model to filter the DILI literature from irrelevant literature based on the title and abstract of publications with ensemble learning and multiple text vectorization algorithms in NLP. The model showed high classification performance and interpretable results.
	
	
	\section{Methods}
	
	\subsection{Data Description and Data Pre-processing}
	The dataset for modelling comprises approximately 14,000 DILI-related papers (which are referred to as positive samples) and approximately 14,000 papers irrelevant to DILI (which are referred to as negative samples). For the contest released by the Annual International Conference on Critical Assessment of Massive Data Analysis (CAMDA 2021), only 50\% of the positive samples (7,177) and negative samples (7,026) was released while the remaining samples were held out for model assessment. For the hold-out test data, which we refer to as the hold-out test dataset 1 in this study, there are 14,211 samples with labels masked to test the model performance on unseen data. For each sample, there are the publication titles and/or the abstracts. The challenge also releases an additional hold-out test data set with 2,000 abstracts, which we refer to as the hold-out test dataset 2 in this study. 
	
	We partitioned the published data (excluding the hold-out datasets) into 80\% training and 20\% validation data. We pre-processed the free text by lowercasing, removing punctuation, numeric, special characters, multiple white spaces, stop words, and finally tokenized the text with Gensim library on Python 3.7 \cite{rehurek_software_2010}. We also performed stemming by changing the terms into their word stems (e.g. “hepatotoxicity” to “hepatotox”) to reduce the number of distinct terms for sake of avoiding model overfitting. We regarded stemming as a model hyperparameter tuned based on the performance on the validation set.
	
	\subsection{Text Vectorization}
	
	To extract features from the free-text literature, we have used several different text vectorization algorithms to transfer the text into numerical features (i.e. word/sentence vectors): Bag-of-words (BOW), term frequency-inverse document frequency (TF-IDF), word2vec (W2V), and sent2vec (S2V).
	
	Both BOW \cite{harris_distributional_1954} and TF-IDF \cite{sparck_jones_statistical_1972} are based on word counts. They are among the simple word vectorization algorithms which are widely used to classify text. As a basic word vectorization approach, BOW enumerates the number of each term's occurrences in a piece of text and uses the number of occurrences of each term as the feature. Since the BOW relies on the counts of all terms, the dimensionality of the features from the extracted text equals the number of all different terms in the training corpus. Based on the basic features extracted based on BOW, TF-IDF further regularized the features by calculating the ratio of term frequency (TF), which denotes the number of term occurrences, and inverse document frequency (IDF), which denotes the number of text samples that contain this term. As a result, the value of a feature for a sample increases as the number of occurrences in the text increases but decreases as the total number of texts that include the term increases. With the regularization, the TF-IDF algorithms put more emphasis on the rare terms over the entire training corpus. In this study, after we applyied the BOW and TF-IDF word vectorization algorithms, the feature dimensions were 30,753 and 42,452 with/without stemming.
	
	Word2vec (W2V) \cite{mikolov_distributed_2013} is a neural-network-based vectorization algorithm. Without directly relying on the number of occurrences of distinct terms, W2V tries to create an embedding matrix $E$ of the term embeddings with a pre-set for all the terms that occurred in the training corpus. Then, W2V optimizes the embedding matrix $E$ and finally when applying the embedding matrix for downstream tasks, W2V maps the terms to their associated embedding vectors in the embedding matrix. The embedding matrix is the goal of optimization in the W2V training process. The randomly initialized embedding matrix is learned with shallow neural networks based on simple prediction tasks, such as the continuous bag-of-words (CBOW) and the continuous skip-gram. In this study, we used two pre-trained biomedical W2V models:  one was trained on a corpus from Wikipedia, PubMed, and PMC (W2V1) \cite{Pyysalo2013DistributionalSR}, and the other was trained on a corpus from PubMed and MIMIC-III (W2V2) \cite{zhang_biowordvec_2019}. Both models include 16,545,452 terms with an embedding dimension of 200. After converting each term in a text into a 200-dimension embedding, an average of all the term embeddings was taken as the embedding for a text.
	
	S2V is another unsupervised sentence embedding algorithm that allows researchers to compose sentence embeddings using word vectors along with n-gram embeddings \cite{pagliardini_unsupervised_2018}. It simultaneously trains the composition and the embedding vectors. S2V can be regarded as an extension of the word contexts from CBOW to a larger sentence context, while the sentence words are specifically optimized via an unsupervised objective function. In this study, we applied a biomedical S2V model trained on PubMed and MIMIC-III corpora with a dimensionality of 700 for a text \cite{chen_biosentvec_2019}.
	
	\subsection{Classification Model Development and Assessment}
	In the development of the DILI literature classification model, we adopted two protocols: 1) non-ensemble learning, where we developed classifiers based on 80\% of the training data and the four different vectorization algorithms; 2) ensemble learning, where we further separated the 80\% training data into 60\% data for training separate learners and 20\% data for training the meta-learner. The separate learners output the predicted probabilities on the samples for meta-learner training while the meta-learner aggregates the predicted probabilities. The rationale for developing ensemble learning models is because the different embedding algorithms contain much diversity as they adopted different frameworks to compute the word vectors and document vectors. The diversity of the separate learners may aid in boosting the performance of document classification via the fusion of predicted probabilities given by different models. It is worth noting that, in the ensemble learning protocol, we discarded the BOW model because it generally performs inferiorly compared with TF-IDF while being similar to TF-IDF as word-count-based algorithms \cite{maaten_visualizing_2008}. Additionally, we added three different weights to the positive/negative classes for the TF-IDF/W2V/S2V models in the ensemble learning, to add divergence and focus more on the positive cases, because in a real-world application setting, considering the broad range of research fields archived in PubMed, the positive case prevalence is likely to be much lower.
	
	In the development of separate learners and in the non-ensemble learning protocol, we adopted the logistic regression (LR) and random forest (RF) models as the classification algorithms. These two algorithms were applied because of their interpretability of important features in the decision-making process. In the meta-learner training, we used LR to reduce the variance and avoid overfitting caused by more flexible classifiers such as RF and support vector machine (SVM). The hyperparameters including the strength of L2 penalty, different class weights for LR, the number of estimators, the number of maximum splits for RF, were tuned via five-fold cross-validation on the corresponding part of training data, with classification accuracy as the optimization goal.
	
	To evaluate the model performance, on the 20\% validation data partitioned from the released dataset (for which we know the labels), we calculated the classification accuracy, the area under the receiver operating characteristic curve (AUROC), the area under the precision-recall curve (AUPRC), and the F1 score. AUROC is the area under the curve in which the x-axis denotes the false positive rate and the y-axis denotes the true positive rate (TPR), while AUPRC is the area under the curve in which the x-axis denotes the recall and the y-axis denotes the precision. The reasons why we adopted both AUROC and AUPRC in evaluating our models are: firstly, AUROC has been a widely used metric in evaluating binary classifiers without reliance on the decision threshold set on predicted class probability; secondly, AUPRC was also used in this study because it is more sensitive to the prevalence and can better reflect model performance in an imbalanced data set \cite{davis_relationship_2006}. Therefore, AUPRC and F1-score were particularly emphasized as in real-world applications, the prevalence of the DILI-positive cases can be much lower and the AUPRC and F1 score can be more comprehensive in evaluating model performance with different weights on positive/negative cases. For the hold-out validation dataset 1 and dataset 2, the accuracy, F1 score, precision, and recall were calculated to validate the model performance on unseen data.
	
	\section{Results}
	
	\subsection{Data Visualization}
	First, we visualized the text vectors given by TF-IDF and S2V as examples, with the unsupervised non-linear dimensionality reduction method: t-distributed stochastic neighbour embedding (t-SNE), which has been shown effective in visualizing high-dimensional data \cite{maaten_visualizing_2008,liu2021boost}. The results show that the positive samples and the negative samples cluster separately indicating the potential feasibility of classifying the DILI-positive samples (Fig. 1). It should be noted that the t-SNE visualization is completely unsupervised and the clusters shown are labelled with the ground truth labels from the dataset.

	\begin{figure}[!t]
		\centering
		\includegraphics[width=0.95\linewidth]{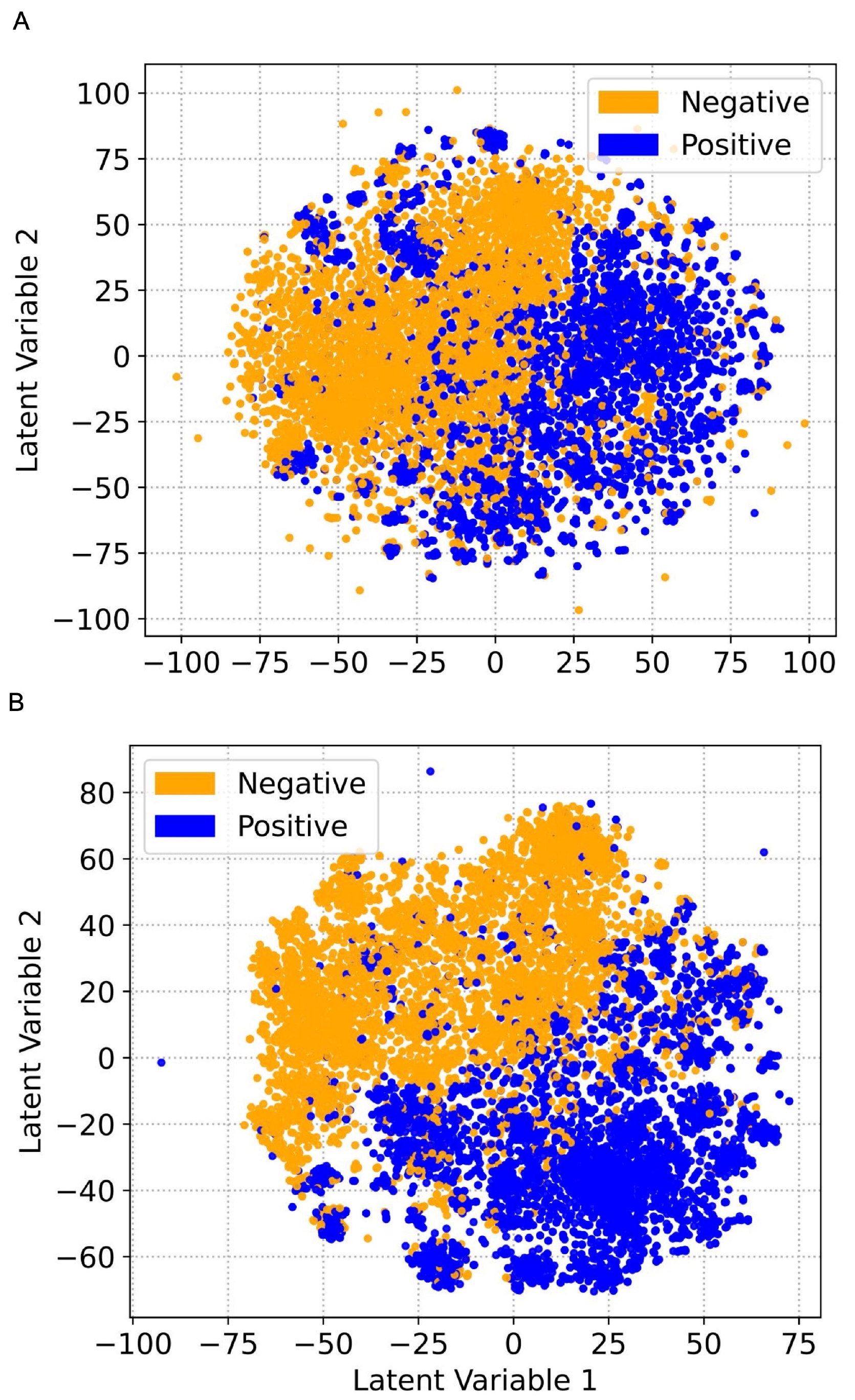}
		\caption{The t-SNE visualization of the text vectors of the training and validation data. (A) TF-IDF text vector visualization; (B) S2V text vector visualization.}
		\label{fig_1}
	\end{figure}
	
	\begin{figure*}[!htp]
		\centering
		\includegraphics[width=0.95\linewidth]{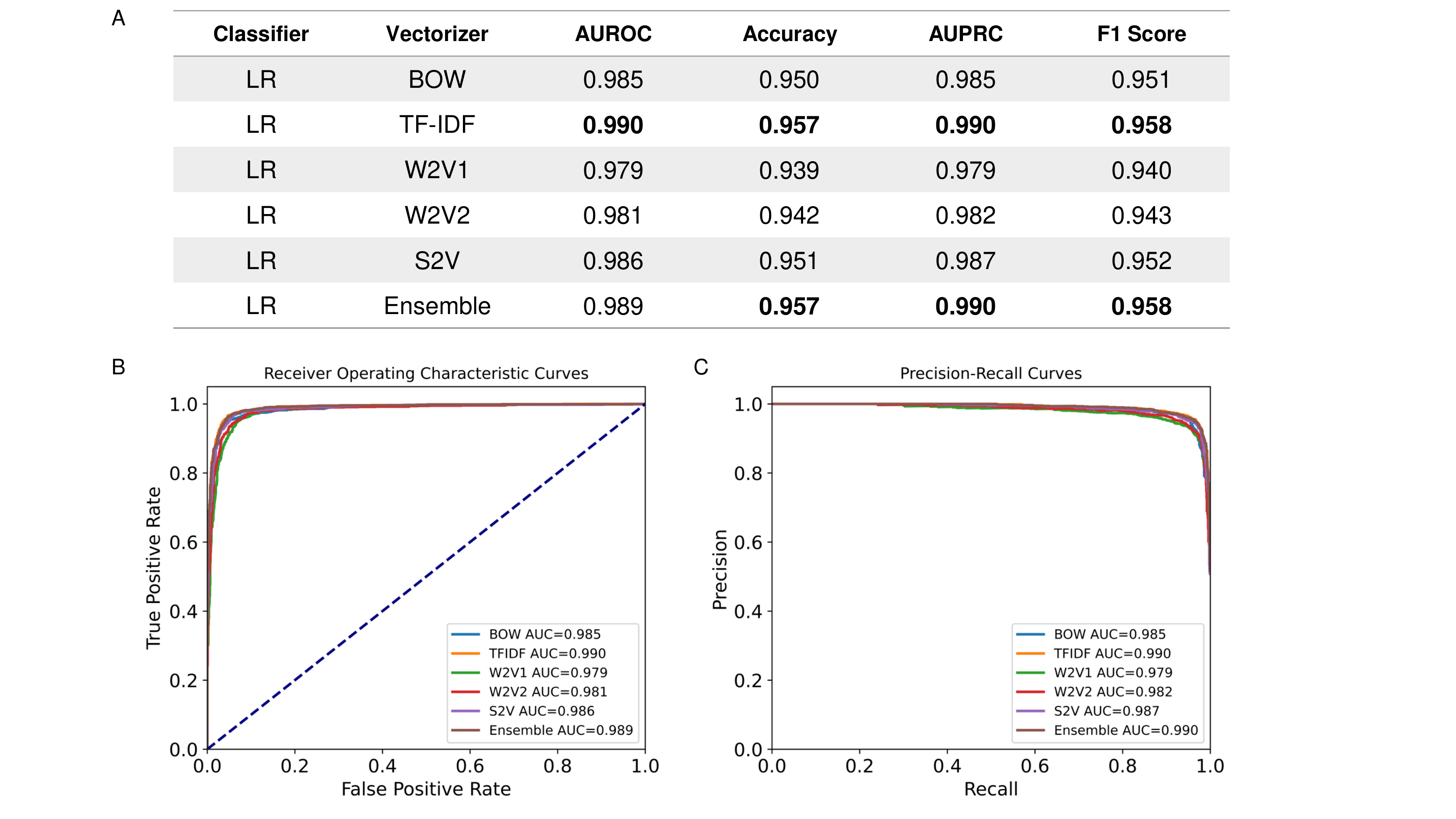}
		\caption{The classification performance of the models. (A) The table of classification performance metrics of different vectorization models.  The receiver operating characteristic curves (B), and the precision-recall curves (C) of the different models.}
		\label{fig_2}
	\end{figure*}

	\subsection{Model Performance on Validation Data}
	
	With text vectors, we built classifiers based on each of the four vectorization algorithms. The results show that besides the ensemble learning model, TF-IDF outperformed the other models with the highest AUROC (0.990), accuracy (0.957), AUPRC (0.990), and F1-score (0.958) (Fig. 2). The RF models did not outperform the LR models and were therefore not shown and used in our ensemble models.

	\begin{figure*}[!htp]
		\centering
		\includegraphics[width=0.9\linewidth]{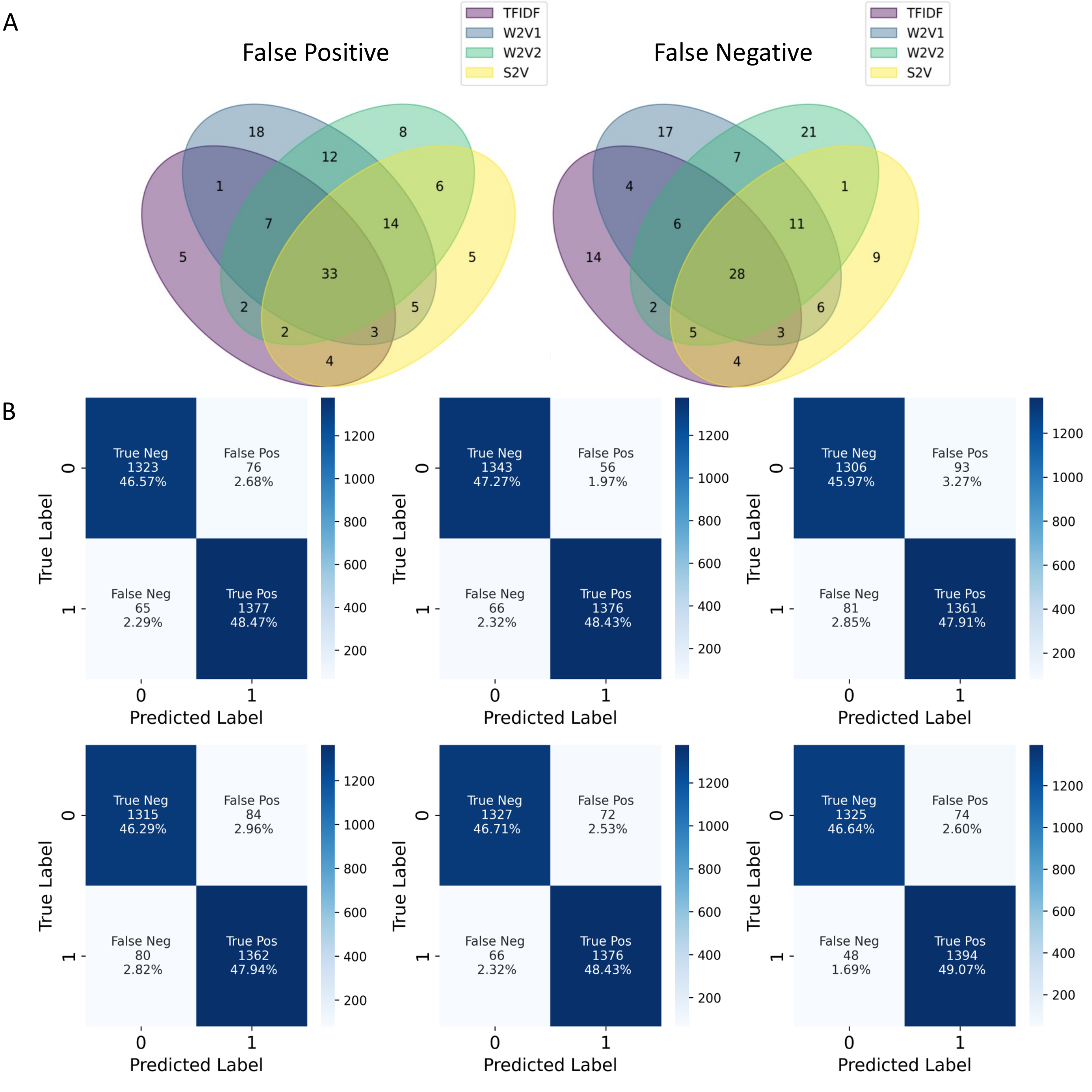}
		\caption{The confusion matrices and the numbers of false-positive cases and false-negative cases of different models. (A) The Venn plot of the false-positive cases and false-negative cases given by four separate learners. (B) The confusion matrices of the six different algorithms, top, from left to right: BOW, TF-IDF, W2V1, bottom: from left to right, W2V2, S2V, and ensemble learning models.}
		\label{fig_3}
	\end{figure*}
	Additionally, after plotting the confusion matrices, it can be seen that among the separate learners, the TF-IDF model has the fewest false-positive cases while the S2V model has the fewest false-negative cases (Fig. 3). We plotted the Venn plots of the false predictions (Fig. 3A). According to the results, although most false-positive and false-negative cases overlap across different word-vectorization models, there is divergence among different models, which motivates the ensemble learning protocol which aggregates the diverse knowledge of diverse learners for potentially better model performance.
	
	We then built the ensemble classifier based on TF-IDF, two W2V models, and S2V. The results show that it has reached the same highest AUPRC, accuracy, and F1-score as the TF-IDF model, with a similar AUROC (Fig. 2). According to the confusion matrices (Fig. 3), the ensemble model shows the fewest false-negative cases and a decent number of false-positive cases. Based on the hyperparameter tuning processes, the hyperparameters of the final ensemble model has the strength of L2 penalty at 10 and the class weight at 1:1. The hyperparameters of the 12 separate LR models leveraged in the ensemble models: strength of L2 penalty: 0.1 for the nine TF-IDF/W2V1/W2V2 models, 1 for three S2V models.
	
	\subsection{Model Interpretation}
	On the validation data, we interpreted the best-performing models of TF-IDF and ensemble learning. For the TF-IDF model, we bootstrapped the training data 1000 times, extracted the coefficients of LR, and averaged the coefficients which correspond to each term. We then ranked the mean coefficients and selected the top 10 most important words for the positive prediction and negative prediction respectively. The results show the important words in the stemmed version: the important words for the positive prediction such as “safety”, “hepatotoxic/hepatotoxicity”, and “liver” show clear meaning related to DILI, illustrating model interpretability (Fig. 4A). Additionally, we visualized the contribution of different vectorization models in the ensemble learning in a similar manner of bootstrapping on the meta-learner training data and finding the normalized LR coefficients (Fig. 4B). The results show that all coefficients are positive indicating the positive contributions of different text vectorization models. Based on these results, it is shown that among these models, TF-IDF models and S2V models perform the best.
	\begin{figure*}[!t]
		\centering
		\includegraphics[width=0.95\linewidth]{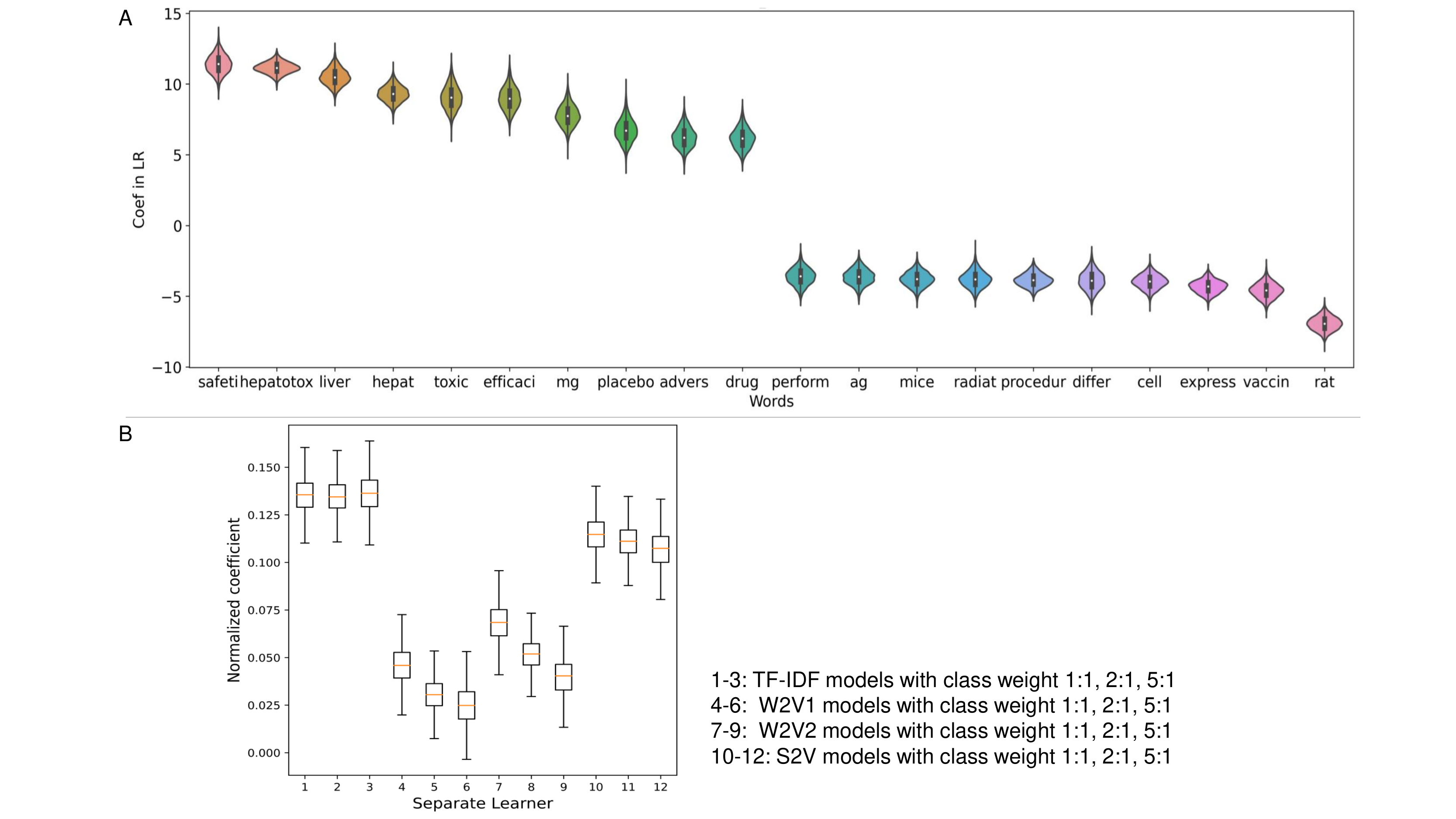}
		\caption{Interpretation of the important words for classification based on TF-IDF and the contribution of separate learners in the ensemble learning model. (A) The top 10 most important words for positive predictions and negative predictions with the distribution of logistic regression coefficients in 2,000 bootstrapping experiments. (B) The normalized logistic regression coefficients of separate learners in 2,000 bootstrapping experiments.}
		\label{fig_4}
	\end{figure*}
	
	\subsection{Model Performance on Hold-out Data and Additional Hold-out Test Data}
	After testing the models on the validation data, we picked up the two best models: the TF-IDF model and the ensemble model and tested them on the hold-out test data. The performance on the hold-out test data is shown in Table 1 and it should be noted that the labels on the hold-out datasets are completely blinded and only the performance metrics can be given upon the submission of predictions. On the hold-out test dataset 1, both models reach the same accuracy of 0.954, while the ensemble model reaches a higher precision of 0.960, and the TF-IDF model reaches a higher recall of 0.961. On the additional hold-out test data (hold-out test dataset 2) with abstract only, the TF-IDF model outperformed the ensemble model with higher accuracy of 0.927, a higher F1 score of 0.930 and a higher precision of 0.886, while the ensemble showed a higher recall of 0.988. Generally, on the two datasets of the hold-out data released by the challenge, both models show high performance in classifying the literature with high accuracy and F1 score, which may enable researchers to accurately filter the DILI-negative literature for further analysis.
	
	\begin{table}[!t]
		\caption{The performance of the  TF-IDF model and the ensemble model on the hold-out data (1) and additional hold-out data (2)}
		\label{table_example}
		\centering
		\begin{tabular}{lllll}
			\hline
			Model      & Accuracy & F1 Score & Precision & Recall \\
			\hline
			TF-IDF-1   & 0.954    & 0.954    & 0.947     & 0.961  \\
			Ensemble-1 & 0.954    & 0.955    & 0.960     & 0.950  \\
			TF-IDF-2   & 0.927    & 0.930    & 0.886     & 0.979  \\
			Ensemble-2 & 0.900    & 0.908    & 0.840     & 0.988 \\
			\hline
		\end{tabular}
	\end{table}

	\section{Discussion}
	In this study, to develop a machine-learning-based model to automatically filter drug-induced liver injury (DILI) related publications out of the irrelevant publications, four different natural language processing (NLP) text vectorization methods were used to vectorize scientific publications. Then, logistic regression models and random forest models were built based on the text vectors and ensemble learning, to predict whether the publications are related to DILI or not. The TF-IDF model and the ensemble learning model with logistic regression classifiers reached the highest classification performance in terms of accuracy, AUPRC, and F1 score on the validation set. Both models show high classification performance on the hold-out data. Additionally, the TF-IDF model is also interpretable with the important words for making positive predictions showing meanings clearly related to DILI, and important words for making negative predictions not directly related to DILI. As DILI, which may cause acute liver failure and even death, has become the major killer of prospective new drug candidates and most DILI research results are in the free-text format in scientific publications, the models developed in this study would enable researchers to fast filter the DILI literature without time-consuming manual work. 
	
	The reason for using an ensemble model is that the different text vectorizations are believed to capture different details in the sentence. The concept is validated by the Venn plot in Fig. 3A which showed overlapping false negatives and false positives of predictions given by models built on various text vectorizations. On top of these overlapping wrong predictions, different text vectorizations also owned their distinctive false positives and false negatives. By using a meta-learner in the ensemble learning process to adopt multiple predictions from different models, this ensemble model could potentially compensate for limitations and boost the strength of different text vectorizations in this task. And this is evidenced by its better prediction results on the hold-out data (Table 1). More importantly, on the validation data, the ensemble learning strategy yields fewer false-negative rate (reduced to 1.69\%, while the individual word-vectorization model can only reach 2.29\%). The fewer false negative predictions better help prevent the missing of DILI-information for researchers, which can be critical for the drug candidates.
	
	As for the application of this study, firstly, the models developed in this study can be applied to filter DILI literature for drug discovery researchers from the large corpus of publications and monitor the latest research on DILI, as assistant systems for information retrieval. Secondly, these literature-filtering models can lay a foundation for future quantitative structure-property relationship (QSPR) modelling in drug discovery because the systems can expedite the DILI-labelling process for different drug candidates effectively \cite{tshitoyan_unsupervised_2019,lo_machine_2018,elbadawi_advanced_2020}. Furthermore, this paradigm of literature-filtering system development can be expanded to other fields of biomedical research as well.
	
	Although the models developed in this study have shown good classification performance, there are limitations in this study that can be further addressed. Firstly, in this study, we mainly applied the context-independent word vectorization and document embedding algorithms to get the features for classification modelling. We haven’t tested transformer-based models such as BERT and GPT-2 or the recurrent-neural-network-based models in this study for the following reasons: 1) the current classification performance has been satisfactory on both the validation set and the two hold-out test sets given by the CAMDA challenge; 2) there have been recent reports which show that the transformer-based models did not show significant improvement over conventional models \cite{pagliardini_unsupervised_2018,wieting2015,arora2016}; 3) the large pre-trained models are complex models which entails a large corpus for fine-tuning and large computational resources as well, which can be hard in this setting as there are only thousands of training data in the corpus; 4) the transformer-based and recurrent-neural-network-based models are not easily interpretable when compared with the conventional word-vector-based method with logistic regression and random forest classifiers. In the future, more studies are warranted to try out context-dependent sentence embeddings. For example, including a pre-trained transformer-based model in ensemble learning that may further improve the classification performance.
	
	Secondly, the training corpus is relatively small for the model training when compared with the majority of natural language processing applications. In this study, we have adopted the pre-trained biomedical word/sentence embedding models such as S2V and W2V to address the issue of limited training data. In the future, a larger training corpus can be made with the addition of multiple additional biomedical text corpus (which may be referred to as unlabelled data for semi-supervised learning or data augmentation) to further improve the model performances on real-world applications. These supplementary data are not necessarily directly related to the DILI topic, but this unbalanced training set better represents the actual application scenario where the majority of the publications on PubMed are irrelevant to DILI.
	\section{Conclusion}
	In this study, several models were developed to filter DILI-related literature based on four text vectorization techniques (bag-of-words, TF-IDF, two biomedical word2vec models, biomedical sent2vec model) and ensemble learning. The model with TF-IDF and LR outperformed others with an AUROC of 0.990, an accuracy of 0.957, and an AUPRC of 0.990. An ensemble learning model with overall similar performance but the fewest false-negative cases was developed based on the prediction probability from 12 individual word-vectorization models, which shows the highest accuracy (0.954) and F1 score on the hold-out data (0.955). Both models performed well on the two hold-out test data with each model taking a lead in different classification metrics (the ensemble learning model accuracy: 0.954,  F1-score: 0.955, precision: 0.960, recall: 0.950). The development of both TF-IDF and ensemble models enables the users to apply these two models for applications and the ensemble-learning-based model enables a more efficient literature filter with fewer false negatives for researchers who focus on the DILI in the field of drug discovery.
	\section*{Acknowledgment}
	
	The authors thank the support from the Department of Bioengineering and the Department of Biomedical Data Science, Stanford University. The authors also thank the CAMDA challenge for the DILI literature data.

	\ifCLASSOPTIONcaptionsoff
	\newpage
	\fi

	
	
	\bibliographystyle{IEEEtran}
	\bibliography{reference}
	%

	%

	
	

\end{document}